\documentclass[aps,twocolumn,showpacs, superscriptaddress,amsmath,amssymb]{revtex4-1}
\usepackage{graphicx}
\usepackage{epsfig}
\usepackage{epstopdf}
\usepackage{bm}
\usepackage[usenames,dvipsnames]{color}
\usepackage[breaklinks=true,colorlinks,citecolor=blue,linkcolor=blue,urlcolor=blue]{hyperref}
\usepackage{float}
\usepackage{soul}

\def\etal{{\it et~al.}}

\begin{document}
\title{Spin Selective Evolution of Zhang-Rice State in Binary Transition Metal Oxide}

\author{Asish K. Kundu}
\email{asishkumar2008@gmail.com}
\affiliation{International Center for Theoretical Physics (ICTP), Trieste, 34151, Italy}
\affiliation{Istituto di Struttura della MateriaCNR (ISM-CNR), Trieste 34149, Italy}
\affiliation{Condensed Matter Physics and Materials Science Department, Brookhaven National Laboratory, Upton, New York 11973, USA}
\author{Polina M. Sheverdyaeva}
\author{Paolo Moras}
\affiliation{Istituto di Struttura della MateriaCNR (ISM-CNR), Trieste 34149, Italy}
\author{Krishnakumar S. R. Menon}
\affiliation{Surface Physics and Material Science Division, Saha Institute of Nuclear Physics, HBNI, 1/AF Bidhannagar, Kolkata 700064, India}
\author{Subhasish Mandal}
\email{subhasish.mandal@mail.wvu.edu}
\affiliation{Department of Physics and Astronomy, West Virginia University, Morgantown, WV 26506, United States.}
\author{Carlo Carbone}
\affiliation{Istituto di Struttura della MateriaCNR (ISM-CNR), Trieste 34149, Italy}

\date{\today}

\begin{abstract}

{The Zhang-Rice (ZR) state is a strongly hybridized bound state formed by the transition metal and oxygen atoms. The spin-fluctuations within the ZR state are known to play an important role in high-$T_\mathrm{c}$ superconductivity in cuprates. Here, we employ a combination of angle-resolved photoemission spectroscopy (ARPES), X-ray photoemission spectroscopy (XPS), and {\it ab initio} embedded dynamical mean-field theory (eDMFT) to investigate the influence of magnetic ordering on the spectral characteristics of the valence band and Mn 2$p$ core-level in MnO (001) ultrathin films. Our results demonstrate that a complex spin-selective evolution of Mn 3$d$$-$O 2$p$ hybridization develops due to the long-range antiferromagnetic (AFM) ordering. This hybridization significantly alters the spectral shape and weight of the ZR state. Specifically, in the AFM phase, we observed the sharpening of the ZR state and band folding with the periodicity of the AFM unit cell of MnO(001). We also demonstrated a strong connection between the spectral evolution of the ZR state and the non-local screening channels of the photoexcited core holes. Further, our detailed temperature-dependent study reveals the presence of short-range antiferromagnetic correlations that exist at much higher temperatures than $T_\mathrm{N}$. Such comprehensive studies showing the evolution of the ZR state across the magnetic transitions and its implication to the core-hole screening have never been reported in any 3$d$ binary transition metal oxides.}

\end{abstract}


\maketitle

The 3$d$ binary transition metal oxides (TMOs) have garnered significant research interest due to their unusual insulating and magnetic properties \cite{shen1990photoemission,shen1991electronic,chen2017lattice,mandal2019influence,SM-pnas}, along with their potential for various technological applications \cite{ohta2003fabrication,jung2012stability}. Having unfilled $d$ orbitals, these oxides should be a metal based on the conventional single-particle band theory; however, they are well-known antiferromagnetic insulators with wide charge gaps. Later, it was found that the strong intra-atomic Coulomb interactions between $d$ electrons are responsible for opening up this insulating charge gap and they can be described by Mott-Hubbard theory \cite{hubbard1964electron,rodl2009quasiparticle}. These TMOs have broad similarities with the high-$T_c$ cuprate superconductors (SC): in both cases, the transition metal atoms are octahedrally coordinated by oxygen atoms, and the parent compound of cuprates and binary TMOs exhibits an antiferromagnetic insulating ground state. Besides, these TMOs host intriguing many-body electronic states such as Zhang-Rice bound state (ZRBS) \cite{zhang1988effective,kunevs2007nio,bala1994zhang}, analogous to the ZR singlet observed in high-$T_c$ cuprate superconductors (SC) \cite{damascelli2003angle,zhang1988effective}. According to the originally proposed model by Zhang and Rice \cite{zhang1988effective}, in hole-doped cuprates, the hybridization strongly binds a hole on each square of oxygen atoms to the central Cu$^{2+}$ ion in the CuO$_2$ plane, and forms a singlet which is known as ZR singlet. It is widely accepted that the ZR singlet in cuprate plays an important role in superconductivity \cite{damascelli2003angle,zhang1988effective,monney2016probing}. Owing to the various similarities between the TMOs and cuprates, understanding the ZR physics in relatively simpler binary TMO systems could help in better understanding the more complex physics of high-$T_\mathrm{c}$ superconducting cuprates.

However, the role of magnetic interactions on the electronic structure and ZRBS of these binary oxides remains controversial due to the conflicting theoretical and experimental results \cite{shen1990photoemission, barman,hermsmeier1990spin,kundu2017effects}. From the experimental side, it is partially due to the difficulties in obtaining high-quality bulk single crystals and the technical challenges of performing low-temperature photoemission measurements due to their insulating nature, which causes potential charging issues. Recent studies have shown that the charging problem can be overcome by growing ultrathin films of these materials on metallic substrates \cite{barman2018growth,kundu2017effects,kundu2016growth,barman,das2015revisit}. However, the high-resolution ARPES studies across the magnetic transition are still spare, with conflicting results \cite{shen1990photoemission, barman,hermsmeier1990spin,kundu2017effects}. For example, in the case of CoO, Shen \etal~ \cite{shen1990photoemission} reported that the ZRBS remain unchanged across the magnetic transition, which contradicts the recent ARPES studies by Barman \etal~\cite{barman}, where strong sharpening of ZRBS is observed in the AFM phase, in line with theoretical predictions \cite{terakura1984band,mandal2019influence}. Similar conflicting results are also found for MnO \cite{hermsmeier1990spin, kundu2017effects}. In the case of NiO, most of the ARPES studies were only performed in the AFM phase \cite{shen1991electronic,shen1990aspects}. Accessing the PM phase requires annealing the sample above $T_N$ = 525 K at which sample decomposition occurs which limits the reliable measurements across $T_N$ \cite{shen1991electronic}.

Theoretical modeling of the electronic structure of these compounds is further challenging due to simultaneously describing both localized and itinerant pictures of electrons. Most of the first-principles approach based on density functional theory (DFT) and various beyond-DFT methods~\cite{mandal2019systematic} such as DFT+$U$, hybrid functionals, and GW-approximation face a great challenge to properly describe the fluctuating moments in time and fail to open up an insulating gap in the paramagnetic phase of the TMOs. The exception is dynamical mean field theory (DMFT) in combination with DFT, where one can obtain proper PM and AFM phases without broken-symmetry configuration {\it a priori}, has become a gold standard for investigating correlated materials~\cite{mandal2019systematic,haule3,Haule_prb10,DMFT_review,Kunes:2008bh,FeSe_monolayer,Mandal:2014,Mandal2:2014,Mandal:2018}. Interestingly, recent DFT+DMFT studies predicted a strong connection between the evolution of ZRBS with magnetic ordering and non-local screening channels of photo-excited core holes \cite{hariki2013dynamical,hariki2017lda}. In contrast, the measured core-level photoemission spectra of CoO do not exhibit any measurable changes across $T_N$  \cite{shen1990photoemission}. All these conflicting results highlight the need for further combined experimental and theoretical studies for an improved understanding of the electronic structure of these materials.

\begin{figure*}[ht!]
\centering
\includegraphics[width=16cm]{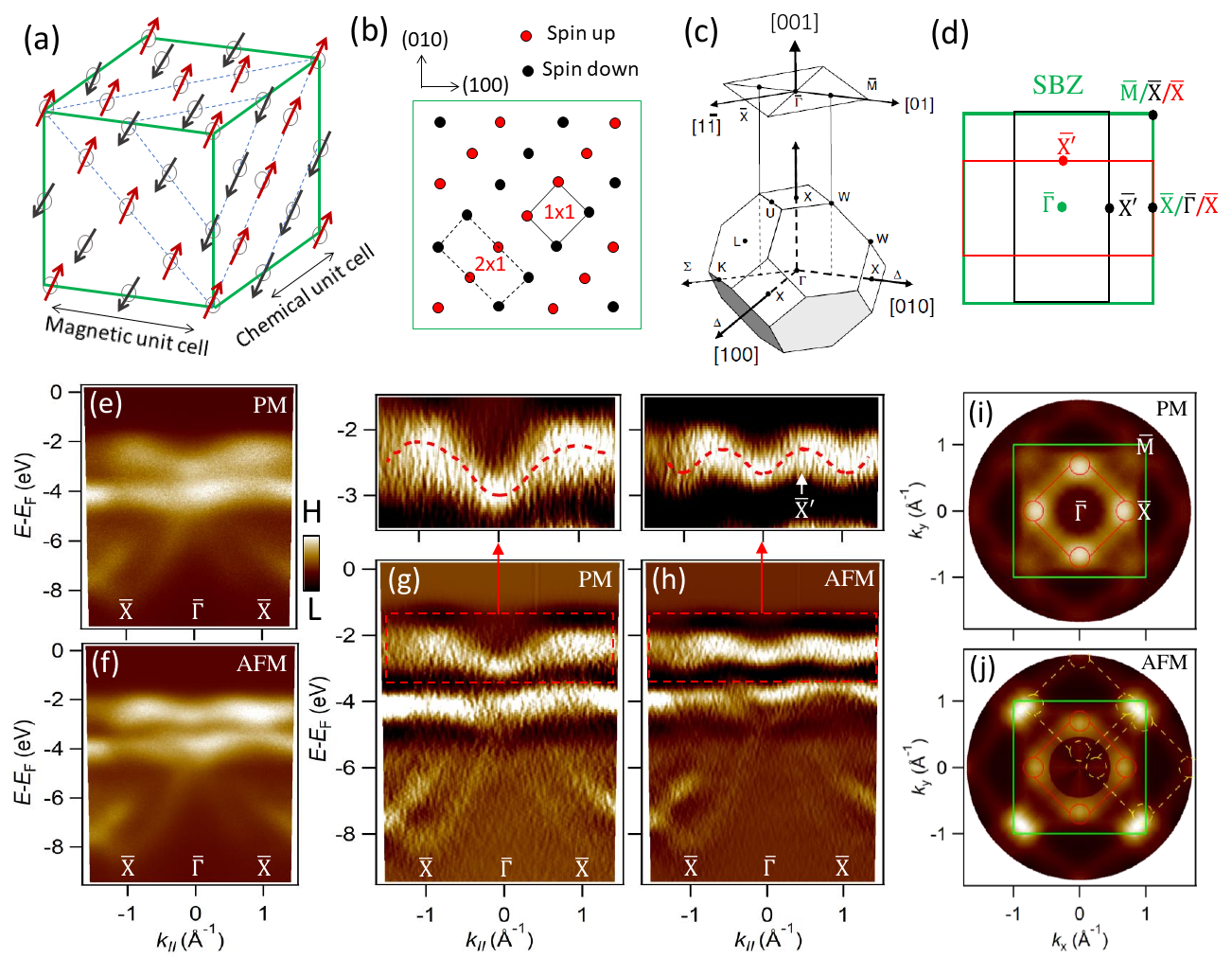}
\caption{Crystal structure and electronic band dispersion of MnO films on Ag(001) showing magnetism-induced band folding. (a) Schematic diagram of the MnO lattice, including spin arrangements of Mn$^{2+}$. The O$^{2-}$ ions are not shown in the figure. The red and black arrows represent the antiparallel arrangement of spins between two consecutive (111) planes. (b) The spin arrangements of Mn$^{2+}$ ions on the MnO(100) surface. The dashed (1$\times$1) square and (2$\times$1) rectangle represent the chemical and magnetic unit cells of MnO(001), respectively. (c) Bulk Brillouin zone (BZ) of MnO and its [001] surface projection. (d) The schematic of the (1$\times$1) SBZ (green) and two orthogonal (2$\times$1) antiferromagnetic domains (red and black). (e) and (f), Energy-momentum dispersions along $\bar{X}$-$\bar{\Gamma}$-$\bar{X}$ path of the MnO(001) SBZ using a photon energy of 125 eV ($k_z$ = 0) at 300 K and 20 K, respectively. (g) and (h), The second derivatives of (e) and (f), respectively. The inset shows the zoomed view of the top part of the valence band (Zhang-Rice bound state, ZRB) as indicated by dashed rectangles in (e) and (f). Dashed curves (red) are used to guide the eye showing the band folding. (i) and (j), Constant-energy contours through the ZRB state in the PM and AFM phase, respectively using a photon energy of 40 eV. The square (green) denotes the SBZ. Solid (red) and dashed (yellow) squares with circles at the corners denote the spectral features in the Fermi surface and their folding, respectively.}\label{Fig1}
\end{figure*}

To address these issues, we chose MnO(001) thin film as a prototype system and performed a detailed electronic structure study across $T_N$ using ARPES and XPS techniques and complemented with our embedded-DMFT (eDMFT) calculations. MnO is chosen as Mn$^{2+}$ has the highest spin state of 5/2 among the transition-metal monoxides and is therefore expected to exhibit significant changes in electronic structure across its N{\'e}el temperature ($T_N \sim$ 120 K) due to strong fluctuating moments. ARPES results show band folding and sharpening of ZRBS across $T_N$. Theoretical results demonstrate that the sharpening of ZRBS in the AFM state is directly linked to the strongly enhanced hybridization between the minority spin channel of Mn e$_g$ and O 2$p$. We have also shown that the change in the hybridization of ZRBS across $T_N$ is strongly connected to the non-local screening channel of Mn 2$p$ core-hole. Details about the experimental and theoretical methods can be found in the Supplemental Material (SM).

\begin{figure*}[ht!]
\centering
\includegraphics[width=16cm]{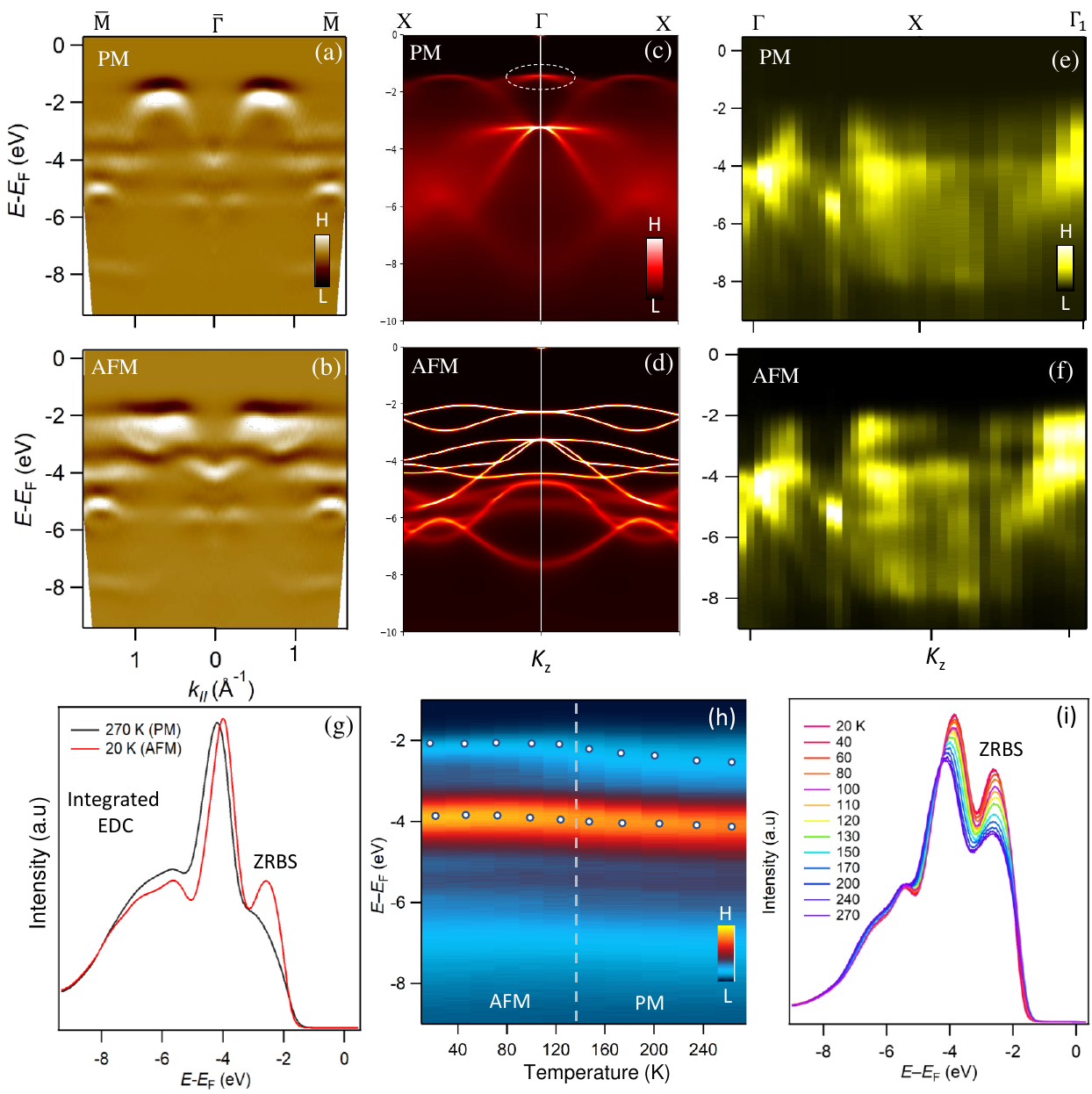}
\caption {ARPES and eDMFT spectra of MnO across $T_\mathrm{N}$. (a) and (b), ARPES spectra of MnO (001) films along the $\bar{M}$-$\bar{\Gamma}$-$\bar{M}$ path of the SBZ using a photon energy of 40 eV at 300 K and 20 K, respectively. (c) and (d), Theoretical band structure using eDMFT along X-$\Gamma$-X path of the bulk BZ in the PM and AFM phase. States enclosed by an ellipse (dashed lines) in (c) appear due to the band folding because of the large PM unit cell taken in our calculations. The $k_z$ dependency of electronic band structure using ARPES along the $\Gamma$-$X$-$\Gamma_1$ directions bulk BZ, perpendicular to the (001): (e) PM ($h\nu=$ 25$-$125 eV, 300 K) and (f) AFM ($h\nu= $25$-$140 eV, 20 K) phase, respectively. (g) Integrated EDCs within the whole momentum range ($\Gamma$-$X$-$\Gamma_1$) from (e) and (f). (h) Stack of normal emission ARPES spectra ($\bar{\Gamma}$ = 0) for various temperatures showing the ZRBS evolution (the data were collected at $h\nu= $ 40 eV). The vertical dashed line indicates the $T_N$ in our MnO(001) film. Filled circles outline the dispersion of the electronic states. (i) EDCs at various temperatures integrated over $\bar{\Gamma} \pm$ 0.5 \AA$^{-1}$ along $\bar{\Gamma}-\bar{X}$.}\label{Fig2}
\end{figure*}

Figure~\ref{Fig1}(a) shows the schematic arrangements of Mn$^{2+}$ ions and their spins in the magnetic unit cell of MnO. The magnetic moments of Mn$^{2+}$ ions are aligned parallel within the (111) planes but antiparallel between adjacent (111) planes \cite{blech1966long}. This type of spin orientation naturally produces in-plane AFM order on the surface of MnO(001) films with $p$(2$\times$1) translation symmetry w.r.t the chemical unit cell of the MnO(001), illustrated in Fig.~\ref{Fig1}(b). The Bulk Brillouin zone (BZ) of MnO and its [001] surface projection are shown in Fig.~\ref{Fig1}(c). The symmetry points without a bar are used to represent the bulk phase, while those with a bar are used to represent the surface projection. From Fig.~\ref{Fig1}(c) it can be seen that the (001) surface projection of $\Gamma$-X direction is the same as the $\bar{\Gamma}$-$\bar{M}$ direction; they are essentially the same at $k_z=$ 0. Further, the schematics of the chemical (green) and magnetic (red and black) surface Brillouin zones (SBZs) are shown in Fig.~\ref{Fig1}(d). Two orthogonal magnetic SBZ are drawn as our MnO(001) films exhibit twin magnetic domains due to the 4-fold rotational symmetry of the Ag(001) substrate \cite{kundu2018evolution}. These AFM domains were resolved in our previous experiments on MnO(001) \cite{kundu2018evolution} and for other binary 3$d$ TMOs \cite{barman,menon2011surface,das2018evolution}.




First, we will discuss the implication of AFM ordering to the electronic structure. In the AFM phase, band folding is expected along the $\bar{\Gamma}$-$\bar{X}$ direction due to the doubling of unit-cell dimension compared to the PM phase. However, here, the situation could be more complex due to the presence of two orthogonal magnetic domains. This is because folding is expected along the $\bar{\Gamma}$-$\bar{X}$ direction for one domain [2$\times$1 (black)], whereas no folding is expected for another domain [1$\times$2 (red)] along the same direction [Fig.~\ref{Fig1}(d)]. Thus, in the AFM phase, we expect a superposition of folded and unfolded bands along $\bar{\Gamma}$-$\bar{X}$. The ARPES spectra along this path are shown for the PM and AFM phases in Fig.~\ref{Fig1}(e) and \ref{Fig1}(f), respectively. Figures~\ref{Fig1}(g) and (h) show their respective second derivatives. Insets show the zoomed view of the topmost valence band. The electronic states in the upper part of the valence band (between $-$1.5 to $-$5 eV) show less dispersion compared to the lower part (between $-$5 to $-$9 eV) as the former region is dominated by strongly correlated Mn 3$d$, while the later by less-correlated O 2$p$ states \cite{lad1988electronic,nekrasov2013consistent,kundu2017effects}. According to our previous eDMFT computations, within the Mn 3$d$ dominated region, the topmost part is mostly due to the e$_g$ character, while the lower part is dominated by t$_{2g}$ character \cite{lad1988electronic,nekrasov2013consistent,eder2008correlated}. However, significant Mn 3$d-$O 2$p$ hybridization has been observed throughout the valence band \cite{nekrasov2013consistent,eder2008correlated,mandal2019influence,trimarchi2018polymorphous}. The topmost valence band is a hybridized bound state formed by the Mn e$_g$ and O 2$p$ orbitals, which is called ZRBS. From our ARPES data, it can be seen that the dispersion of this state changes between the PM to AFM phases and it shows folding in the AFM phase, with a periodicity two times higher than that of the PM phase [inset of Figs.~\ref{Fig1}(g) and (h)]. The effect of band folding can also be seen from the constant energy cuts (at $E=-$2.1 eV) as shown in Figs.~\ref{Fig1}(i) and (j). In the PM phase, high-intense spots are observed near the $\bar{X}$ whereas, in the case of the AFM phase, additional high-intense spots appear close to the $\bar{M}$. It is exactly what is expected from the band folding scenario as $\bar{M}$ (for the 1$\times$1 cell) becomes $\bar{X}$ (for 2$\times$1 and 1$\times$2 cells), thus the spectral features near $\bar{X}$ should be reflected around $\bar{M}$. The residual intensity observed around the $\bar{X}$ in the AFM phase can be attributed to the simultaneous presence of both folded and unfolded bands as we discussed earlier. Between the PM and AFM phases, some changes in band dispersion are also seen for the second valence band (around $-$4 eV), however, no measurable change has been observed for the states with dominant O 2$p$ characters. Similar behavior was observed for the antiferromagnetic MnO$_2$ chains \cite{schmitt2019indirect}. This could be due to the fact that the magnetic moments are sitting on the Mn$^{2+}$ sites not on O$^{2-}$. Thus the Mn 3$d$ electrons could dominantly feel the interactions with the spin compared to the O$^{2-}$ electrons.

\begin{figure*}[ht!]
\centering
\includegraphics[width=17cm]{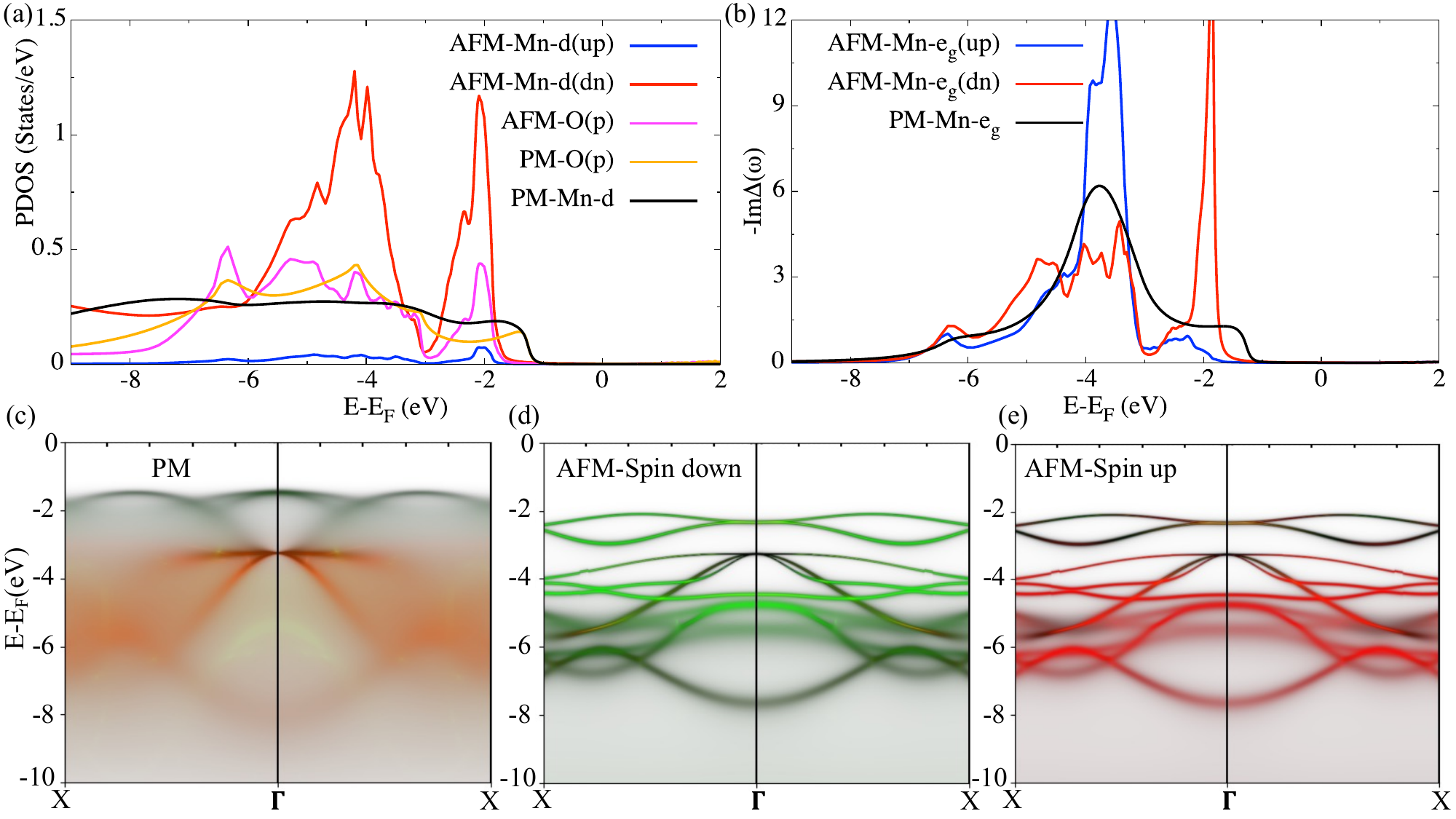}
\caption { Partial density of states (PDOS) and hybridization function computed using eDMFT in the PM and AFM phase. (a) PDOS of Mn $3d$ and O 2$p$. (b) Hybridization function. (c) Orbital ( Mn $3d$ and O 2$p$) projected band structure for the  (c) PM and (d, e) AFM spin-down and spin-up, respectively. Green and red represent the Mn $3d$ and O 2$p$ contributions, respectively; red color is enhanced to show  O 2$p$ clearly.}\label{Fig3}
\end{figure*}

Across the magnetic transition, a distinct change in the band dispersion of the ZRBS is also observed along the $\bar{\Gamma}$-$\bar{M}$ direction, illustrated in Figs.~\ref{Fig2}(a) and (b). Especially, in the AFM phase, the two branches of the ZRBS are observed at around the midway between $\bar{\Gamma}$-$\bar{M}$ [Fig.~\ref{Fig2}(b)], and their dispersions are well reproduced by our eDMFT calculations [Fig.~\ref{Fig2}(d)]. However, in the PM phase, an apparent discrepancy in the dispersion of ZRBS can be seen between the theory and experiment around $\bar{\Gamma}$, where an extra band (enclosed by a dotted ellipse) is seen around $-$2 eV in theory [Fig.~\ref{Fig2}(c)], which is absent in the experiment. Further, eDMFT computations reveal that this extra band appears due to the folding of the electronic state because of the large dimension of the PM cell (AFM cell with random spins) considered in our calculations; this band disappears if we perform eDMFT computation in a PM cell (see also Fig. S1 in the SM for comparison). To exactly probe the electronic structure along the X-$\Gamma$-X path, we have performed photon energy dependent ($k_z$ dependence) ARPES measurements in a wide photon energy range that covers two consecutive ${\Gamma}$ points in the bulk BZ, as shown in Figs.~\ref{Fig2}(e) and (f). By comparing ARPES data with the theoretical band dispersion [Fig.~\ref{Fig2}(c)], the symmetry point ${\Gamma}$ is identified, where all the bands apparently touch each other, similar to what is also observed for NiO and CoO \cite{shen1990aspects,shen1990photoemission,nekrasov2013consistent,mandal2019influence}. From the symmetry of the band dispersion, we estimated the value of the crystal potential (V$_0$) to be $\sim$ 6.0 eV. In Figs.~\ref{Fig2}(e) and (f), the ARPES spectra at around ${\Gamma X/2}$ (41 eV$\leq hv \leq$  49 eV) show strong suppression of photoelectron intensity due to the antiresonance \cite{lad1988electronic}. Further, we noticed that the spectral intensity of the ZRBS is strongly suppressed around the $\Gamma$ point but enhanced in the next $\Gamma$ (denoted by $\Gamma_1$), which points to the matrix element effect as its origin. Overall, the electronic states in the AFM phase are found sharper and well-defined compared to the PM phase, in agreement with our theoretical results [Fig.~\ref{Fig2}(d)]. The integrated energy distribution curves (EDCs) within the whole momentum range ($\Gamma$-$X$-$\Gamma_1$) also show the same behavior [Fig.~\ref{Fig2}(g)]. Besides, from Fig.~\ref{Fig2}(g), a clear redistribution of spectral weight can be observed in a wide energy range. Spectral weight renormalization is often observed in various strongly correlated magnetic materials due to the intricate change in hybridization across the transition \cite{kundu2021role,han2023interplay,Topo-SM1}.


Further, to understand the nature of electronic reconstructions and their connection to magnetism, we conducted extensive temperature-dependent ARPES studies over a broad temperature range. Figure~\ref{Fig2}(h) shows the temperature evolution of ARPES spectra at $\bar{\Gamma}$ = 0. Figure~\ref{Fig2}(i) represents the EDCs integrated over a wider momentum range ($\bar{\Gamma} \pm$ 0.5 \AA$^{-1}$) for various temperatures. It can be noticed that the dispersion and the spectral weight of bands gradually change in a wide temperature range, without a sharp jump at $T_N$. However, the changes are more between 100 to 170 K ([Fig.~\ref{Fig2}(h) and (i)]. Spectral weight change at a much higher temperature than $T_N$, suggests the presence of short-range AFM order, which is in line with the neutron diffraction and spin-polarized photoelectron diffraction SPPD experiments reported on bulk MnO single crystals \cite{alan,herm,hermsmeier1990spin}. We note that the ARPES spectra also show relatively less change between the AFM and PM phases compared to our eDMFT calculations. This is possibly due to the presence of short-range AFM ordering even in the PM phase and finite $k_z$ broadening in ARPES.

To get insight into the observed electronic structure reconstructions, we have calculated the partial density of states (DOS), hybridization function, and orbital projected electronic band dispersions in the PM and AFM phase using eDMFT and present them in Fig.~\ref{Fig3}. From Fig.~\ref{Fig3}(a), we notice that the first valence peak, which is the ZRBS, is a hybridized state between minority spin Mn-$d$ and O-$2p$. We also noticed that the ZRBS gets sharpened in the AFM phase compared to the PM phase, consistent with our ARPES results. To better understand this, we show eDMFT computed hybridization functions in Fig.~\ref{Fig3}(b) for both PM and AFM phases for Mn-$e_g$ electrons. It's evident from Fig.~\ref{Fig3}(b) that the peak (in red) in the hybridization function sharpening for the ZRBS in the AFM phase. Interestingly, the sharpening is observed only for the minority spin-channel as the contribution for the majority channel is found to be relatively much smaller for the first peak and grows only around $-$4 eV for the majority spin channel (in blue), where the PM phase also shows a peak (black). Next, we plot the orbital and $k$-resolved eDMFT spectral functions for the PM phase in Fig.~\ref{Fig3}(c). For the AFM phase, we show this for both spin minority or down (Fig.~\ref{Fig3}d) and the majority or up (Fig.~\ref{Fig3}e) components. In the PM phase, we notice that the first valence peak consists of the Mn-$d$ and O-$2p$. For the AFM phase, the orbital contribution in the spectral functions for majority and minority channels are distinctly different. The Mn-$e_g$ electrons mostly contribute to the spectral function for the minority channel (green), while for the majority channel, it is mostly due to O-$2p$ (red), which is consistent with the PDOS.

\begin{figure}
\centering
\includegraphics[width=7.5cm]{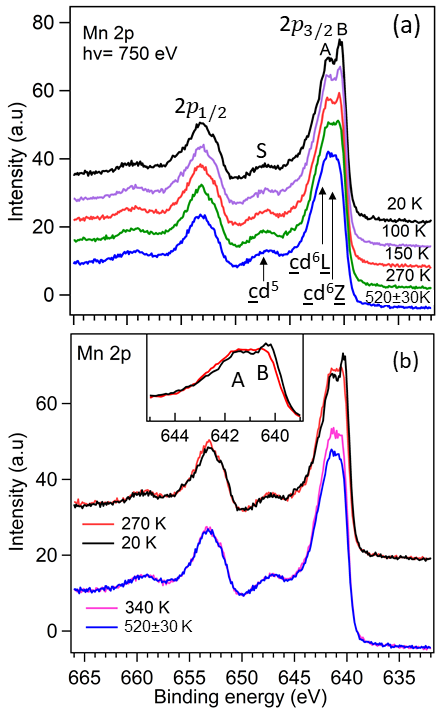}
\caption {Temperature dependence of the Mn 2$p$ core-level of MnO films showing the variation of local- and nonlocal screening of core-hole. (a) Temperature dependence of Mn 2$p$ core-levels using a photon energy of 750 eV, where A and B peaks are the multiplet structure of 2$p_{3/2}$ associated with the local- and nonlocal screening, respectively and S is the charge transfer satellite. States are marked with the final state photoemission as obtained from the Ligand field picture. (b) Same spectra as (a) but superimposed on top of one another to understand the overall change of spectral shape. Inset shows zoomed-view of Mn 2$p_{3/2}$ peak.}\label{Fig4}
\end{figure}

Now we turn our discussions to the core-level spectra. Before discussing our results, it’s important to note that theoretical studies using the DMFT approach on 3$d$ transition-metal oxides have shown that there is a strong connection between the ZRBS hybridization and the multiplet structure of 2$p$ core levels \cite{hariki2013dynamical,hariki2017lda}. It has been shown that, in the AFM phase of NiO, the Zhang-Rice peak gets sharpened compared to the PM phase which leads to the dominance of nonlocal core-hole screening over local screening. Furthermore, the relative strength of these screenings is determined by both the spin arrangement of the crystals and the hybridization function \cite{hariki2013dynamical}. To understand how these effects act on Mn 2$p$ core-levels of MnO, a detailed temperature dependence study has been conducted as shown in Figure~\ref{Fig4}. From Fig.~\ref{Fig4}(a) it can be seen that both Mn 2$p_{1/2}$ and Mn 2$p_{3/2}$ peaks show multiplet structures, whereas it is better resolved for the later and at low temperatures. The features denoted as A, B, and S in the XPS spectra are predominantly originating from the $\underbar{c}d^6\underbar{L}$, $\underbar{c}d^6\underbar{Z}$ and $\underbar{c}d^5$ photoemission final states, respectively, where  $\underbar{c}$, $\underbar{L}$, and $\underbar{Z}$, represent a hole in the Mn 2{\it p}, O 2{\it p} ligand and ZRBS, respectively \cite{hariki2017lda,hariki2013dynamical,horiba2004nature}. That means the peak `A' is associated with the local charge transfer screening of the core-hole within the core-excited MnO$_6$ octahedron while the peak `B' is due to the nonlocal screening (NLS) accompanied by ZRBS \cite{hariki2013dynamical,horiba2004nature}. By lowering temperatures, we noticed a gradual enhancement of the intensity of the B peak relative to the A peak, analogous to the spectral evolution of the ZRBS (Fig.~\ref{Fig2}(a,d)). This suggests nonlocal screening dominates over local screening in the AFM phase compared to PM. Thus, our results clearly represent that there is a strong coupling between the hybridization strength of ZRBS and the screening channels of core holes, in agreement with the theoretical predictions \cite{hariki2013dynamical,hariki2017lda}. Strong dependence of the line shape of Mn 2{\it p} on the magnetic state has been also observed for different manganites \cite{van2006competition,horiba2004nature}.

Another important observation is that within the paramagnetic phase, Mn 2$p_{3/2}$ peak intensity significantly drops with increasing temperature from 340 to 500 K, whereas the intensity of Mn 2$p_{1/2}$ peak is almost constant (Fig.~\ref{Fig4}(b)). As the intensity of these spin-orbit split states is determined by the degeneracy factor (2$J+$1, where $J$ is the total angular momentum ), the relative intensity change could mean that breaking of degeneracy. It should be noted that in the case of transition-metal and rare-earth elements with partially occupied $d$ and $f$ levels, the effective spin value within the $J$ is not solely the spin of the remaining unpaired electron (generated during photoemission) but it often interacts with the spin of unpaired valence electrons. Thus the decrease of Mn 2$p_{3/2}$ intensity at high temperatures could be due to the decrease of the interaction strength between the spin-up component of Mn 2{\it p} and the Mn 3{\it d} spin as short-range magnetic correlation collapses $\sim$ 530 K \cite{herm}. Is it important to note that, a strong dependency of the Mn 3$s$ peak intensity and line shape was previously observed by Hermsmeier {\it et al.} and was explained due to the change of interaction strength between 3$s$ and 3$d$ spins \cite{hermsmeier1990spin}. However, further studies such as spin-polarized XPS are needed to verify the exact origin of the Mn 2$p_{3/2}$ intensity change, which is outside the scope of our present study.

In summary, we have reported detailed electronic structure results of MnO(001) thin films across $T_N$ by ARPES, XPS measurements, and complemented by first-principles eDMFT computations. In spite of the strongly localized character of valence bands of 3{\it d} binary transition metal oxides, here, we clearly resolve band-folding and strong spectral evolution due to the AFM-II spin ordering. An enhancement of ZRBS intensity and the overall spectra are found to sharpen in the AFM phase due to the spin-dependent change in Mn 3$d$$-$O 2$p$ hybridization. We explicitly show that the strength of the hybridization significantly grows in the AFM phase only in the minority spin channel, which is subject to stronger spin-fluctuations. We further show that enhancement of this hybridization strength in ZRBS has a significant effect on the non-local screening channel of 2$p$ core-hole, in agreement with the theoretical predictions. By performing extensive temperature-dependent ARPES and XPS studies we found that the spectral evolution persists at much higher temperatures than $T_N$, which suggests the presence of short-range AFM correlation even at the PM phase. Finally, we believe that our robust observation of spin-dependent change in the valence band and core-level electronic structure can be observed for other similar metal oxides.

A.K.K. acknowledges receipt of a fellowship from the ICTP-TRIL Programme, Trieste, Italy. During the preparation of the manuscript, A.K.K. received funding from the US Department of Energy, Office of Basic Energy Sciences, contract no. DE-SC0012704. S. M. acknowledges the support from the Air Force Office of Scientific Research by the Department of Defense under the award number FA9550-23-1-0498 of the DEPSCoR program and benefited from the Frontera supercomputer at the Texas Advanced Computing Center (TACC) at The University of Texas at Austin, which is supported by National Science Foundation grant number OAC-1818253.


\begin{thebibliography}{49}%
\makeatletter
\providecommand \@ifxundefined [1]{%
 \@ifx{#1\undefined}
}%
\providecommand \@ifnum [1]{%
 \ifnum #1\expandafter \@firstoftwo
 \else \expandafter \@secondoftwo
 \fi
}%
\providecommand \@ifx [1]{%
 \ifx #1\expandafter \@firstoftwo
 \else \expandafter \@secondoftwo
 \fi
}%
\providecommand \natexlab [1]{#1}%
\providecommand \enquote  [1]{``#1''}%
\providecommand \bibnamefont  [1]{#1}%
\providecommand \bibfnamefont [1]{#1}%
\providecommand \citenamefont [1]{#1}%
\providecommand \href@noop [0]{\@secondoftwo}%
\providecommand \href [0]{\begingroup \@sanitize@url \@href}%
\providecommand \@href[1]{\@@startlink{#1}\@@href}%
\providecommand \@@href[1]{\endgroup#1\@@endlink}%
\providecommand \@sanitize@url [0]{\catcode `\\12\catcode `\$12\catcode
  `\&12\catcode `\#12\catcode `\^12\catcode `\_12\catcode `\%12\relax}%
\providecommand \@@startlink[1]{}%
\providecommand \@@endlink[0]{}%
\providecommand \url  [0]{\begingroup\@sanitize@url \@url }%
\providecommand \@url [1]{\endgroup\@href {#1}{\urlprefix }}%
\providecommand \urlprefix  [0]{URL }%
\providecommand \Eprint [0]{\href }%
\providecommand \doibase [0]{http://dx.doi.org/}%
\providecommand \selectlanguage [0]{\@gobble}%
\providecommand \bibinfo  [0]{\@secondoftwo}%
\providecommand \bibfield  [0]{\@secondoftwo}%
\providecommand \translation [1]{[#1]}%
\providecommand \BibitemOpen [0]{}%
\providecommand \bibitemStop [0]{}%
\providecommand \bibitemNoStop [0]{.\EOS\space}%
\providecommand \EOS [0]{\spacefactor3000\relax}%
\providecommand \BibitemShut  [1]{\csname bibitem#1\endcsname}%
\let\auto@bib@innerbib\@empty
\bibitem [{\citenamefont {Shen}\ \emph
  {et~al.}(1990{\natexlab{a}})\citenamefont {Shen}, \citenamefont {Allen},
  \citenamefont {Lindberg}, \citenamefont {Dessau}, \citenamefont {Wells},
  \citenamefont {Borg}, \citenamefont {Ellis}, \citenamefont {Kang},
  \citenamefont {Oh}, \citenamefont {Lindau} \emph
  {et~al.}}]{shen1990photoemission}%
  \BibitemOpen
  \bibfield  {author} {\bibinfo {author} {\bibfnamefont {Z.-X.}\ \bibnamefont
  {Shen}}, \bibinfo {author} {\bibfnamefont {J.}~\bibnamefont {Allen}},
  \bibinfo {author} {\bibfnamefont {P.}~\bibnamefont {Lindberg}}, \bibinfo
  {author} {\bibfnamefont {D.}~\bibnamefont {Dessau}}, \bibinfo {author}
  {\bibfnamefont {B.}~\bibnamefont {Wells}}, \bibinfo {author} {\bibfnamefont
  {A.}~\bibnamefont {Borg}}, \bibinfo {author} {\bibfnamefont {W.}~\bibnamefont
  {Ellis}}, \bibinfo {author} {\bibfnamefont {J.}~\bibnamefont {Kang}},
  \bibinfo {author} {\bibfnamefont {S.-J.}\ \bibnamefont {Oh}}, \bibinfo
  {author} {\bibfnamefont {I.}~\bibnamefont {Lindau}},  \emph {et~al.},\
  }\href@noop {} {\bibfield  {journal} {\bibinfo  {journal} {Phys. Rev. B}\
  }\textbf {\bibinfo {volume} {42}},\ \bibinfo {pages} {1817} (\bibinfo {year}
  {1990}{\natexlab{a}})}\BibitemShut {NoStop}%
\bibitem [{\citenamefont {Shen}\ \emph {et~al.}(1991)\citenamefont {Shen},
  \citenamefont {List}, \citenamefont {Dessau}, \citenamefont {Wells},
  \citenamefont {Jepsen}, \citenamefont {Arko}, \citenamefont {Barttlet},
  \citenamefont {Shih}, \citenamefont {Parmigiani}, \citenamefont {Huang} \emph
  {et~al.}}]{shen1991electronic}%
  \BibitemOpen
  \bibfield  {author} {\bibinfo {author} {\bibfnamefont {Z.-X.}\ \bibnamefont
  {Shen}}, \bibinfo {author} {\bibfnamefont {R.}~\bibnamefont {List}}, \bibinfo
  {author} {\bibfnamefont {D.}~\bibnamefont {Dessau}}, \bibinfo {author}
  {\bibfnamefont {B.}~\bibnamefont {Wells}}, \bibinfo {author} {\bibfnamefont
  {O.}~\bibnamefont {Jepsen}}, \bibinfo {author} {\bibfnamefont
  {A.}~\bibnamefont {Arko}}, \bibinfo {author} {\bibfnamefont {R.}~\bibnamefont
  {Barttlet}}, \bibinfo {author} {\bibfnamefont {C.}~\bibnamefont {Shih}},
  \bibinfo {author} {\bibfnamefont {F.}~\bibnamefont {Parmigiani}}, \bibinfo
  {author} {\bibfnamefont {J.}~\bibnamefont {Huang}},  \emph {et~al.},\
  }\href@noop {} {\bibfield  {journal} {\bibinfo  {journal} {Phys. Rev. B}\
  }\textbf {\bibinfo {volume} {44}},\ \bibinfo {pages} {3604} (\bibinfo {year}
  {1991})}\BibitemShut {NoStop}%
\bibitem [{\citenamefont {Chen}\ \emph {et~al.}(2017)\citenamefont {Chen},
  \citenamefont {Sakata}, \citenamefont {Yamauchi}, \citenamefont {Yang},
  \citenamefont {Kumara}, \citenamefont {Song}, \citenamefont {Palina},
  \citenamefont {Taguchi}, \citenamefont {Ina}, \citenamefont {Katsuya} \emph
  {et~al.}}]{chen2017lattice}%
  \BibitemOpen
  \bibfield  {author} {\bibinfo {author} {\bibfnamefont {Y.}~\bibnamefont
  {Chen}}, \bibinfo {author} {\bibfnamefont {O.}~\bibnamefont {Sakata}},
  \bibinfo {author} {\bibfnamefont {R.}~\bibnamefont {Yamauchi}}, \bibinfo
  {author} {\bibfnamefont {A.}~\bibnamefont {Yang}}, \bibinfo {author}
  {\bibfnamefont {L.~S.~R.}\ \bibnamefont {Kumara}}, \bibinfo {author}
  {\bibfnamefont {C.}~\bibnamefont {Song}}, \bibinfo {author} {\bibfnamefont
  {N.}~\bibnamefont {Palina}}, \bibinfo {author} {\bibfnamefont
  {M.}~\bibnamefont {Taguchi}}, \bibinfo {author} {\bibfnamefont
  {T.}~\bibnamefont {Ina}}, \bibinfo {author} {\bibfnamefont {Y.}~\bibnamefont
  {Katsuya}},  \emph {et~al.},\ }\href@noop {} {\bibfield  {journal} {\bibinfo
  {journal} {Phys. Rev. B}\ }\textbf {\bibinfo {volume} {95}},\ \bibinfo
  {pages} {245301} (\bibinfo {year} {2017})}\BibitemShut {NoStop}%
\bibitem [{\citenamefont {Mandal}\ \emph
  {et~al.}(2019{\natexlab{a}})\citenamefont {Mandal}, \citenamefont {Haule},
  \citenamefont {Rabe},\ and\ \citenamefont
  {Vanderbilt}}]{mandal2019influence}%
  \BibitemOpen
  \bibfield  {author} {\bibinfo {author} {\bibfnamefont {S.}~\bibnamefont
  {Mandal}}, \bibinfo {author} {\bibfnamefont {K.}~\bibnamefont {Haule}},
  \bibinfo {author} {\bibfnamefont {K.~M.}\ \bibnamefont {Rabe}}, \ and\
  \bibinfo {author} {\bibfnamefont {D.}~\bibnamefont {Vanderbilt}},\
  }\href@noop {} {\bibfield  {journal} {\bibinfo  {journal} {Phys. Rev. B}\
  }\textbf {\bibinfo {volume} {100}},\ \bibinfo {pages} {245109} (\bibinfo
  {year} {2019}{\natexlab{a}})}\BibitemShut {NoStop}%
\bibitem [{\citenamefont {Zhang}\ \emph {et~al.}(2021)\citenamefont {Zhang},
  \citenamefont {Mondal}, \citenamefont {Mandal}, \citenamefont {Allred},
  \citenamefont {Aghamiri}, \citenamefont {Fali}, \citenamefont {Zhang},
  \citenamefont {Zhou}, \citenamefont {Cao}, \citenamefont {Rodolakis},
  \citenamefont {McChesney}, \citenamefont {Wang}, \citenamefont {Sun},
  \citenamefont {Abate}, \citenamefont {Roy}, \citenamefont {Rabe},\ and\
  \citenamefont {Ramanathan}}]{SM-pnas}%
  \BibitemOpen
  \bibfield  {author} {\bibinfo {author} {\bibfnamefont {Z.}~\bibnamefont
  {Zhang}}, \bibinfo {author} {\bibfnamefont {S.}~\bibnamefont {Mondal}},
  \bibinfo {author} {\bibfnamefont {S.}~\bibnamefont {Mandal}}, \bibinfo
  {author} {\bibfnamefont {J.~M.}\ \bibnamefont {Allred}}, \bibinfo {author}
  {\bibfnamefont {N.~A.}\ \bibnamefont {Aghamiri}}, \bibinfo {author}
  {\bibfnamefont {A.}~\bibnamefont {Fali}}, \bibinfo {author} {\bibfnamefont
  {Z.}~\bibnamefont {Zhang}}, \bibinfo {author} {\bibfnamefont
  {H.}~\bibnamefont {Zhou}}, \bibinfo {author} {\bibfnamefont {H.}~\bibnamefont
  {Cao}}, \bibinfo {author} {\bibfnamefont {F.}~\bibnamefont {Rodolakis}},
  \bibinfo {author} {\bibfnamefont {J.~L.}\ \bibnamefont {McChesney}}, \bibinfo
  {author} {\bibfnamefont {Q.}~\bibnamefont {Wang}}, \bibinfo {author}
  {\bibfnamefont {Y.}~\bibnamefont {Sun}}, \bibinfo {author} {\bibfnamefont
  {Y.}~\bibnamefont {Abate}}, \bibinfo {author} {\bibfnamefont
  {K.}~\bibnamefont {Roy}}, \bibinfo {author} {\bibfnamefont {K.~M.}\
  \bibnamefont {Rabe}}, \ and\ \bibinfo {author} {\bibfnamefont
  {S.}~\bibnamefont {Ramanathan}},\ }\href {\doibase 10.1073/pnas.2017239118}
  {\bibfield  {journal} {\bibinfo  {journal} {Proceedings of the National
  Academy of Sciences}\ }\textbf {\bibinfo {volume} {118}},\ \bibinfo {pages}
  {e2017239118} (\bibinfo {year} {2021})}\BibitemShut {NoStop}%
\bibitem [{\citenamefont {Ohta}\ \emph {et~al.}(2003)\citenamefont {Ohta},
  \citenamefont {Hirano}, \citenamefont {Nakahara}, \citenamefont {Maruta},
  \citenamefont {Tanabe}, \citenamefont {Kamiya}, \citenamefont {Kamiya},\ and\
  \citenamefont {Hosono}}]{ohta2003fabrication}%
  \BibitemOpen
  \bibfield  {author} {\bibinfo {author} {\bibfnamefont {H.}~\bibnamefont
  {Ohta}}, \bibinfo {author} {\bibfnamefont {M.}~\bibnamefont {Hirano}},
  \bibinfo {author} {\bibfnamefont {K.}~\bibnamefont {Nakahara}}, \bibinfo
  {author} {\bibfnamefont {H.}~\bibnamefont {Maruta}}, \bibinfo {author}
  {\bibfnamefont {T.}~\bibnamefont {Tanabe}}, \bibinfo {author} {\bibfnamefont
  {M.}~\bibnamefont {Kamiya}}, \bibinfo {author} {\bibfnamefont
  {T.}~\bibnamefont {Kamiya}}, \ and\ \bibinfo {author} {\bibfnamefont
  {H.}~\bibnamefont {Hosono}},\ }\href@noop {} {\bibfield  {journal} {\bibinfo
  {journal} {Applied physics letters}\ }\textbf {\bibinfo {volume} {83}},\
  \bibinfo {pages} {1029} (\bibinfo {year} {2003})}\BibitemShut {NoStop}%
\bibitem [{\citenamefont {Jung}\ \emph {et~al.}(2012)\citenamefont {Jung},
  \citenamefont {Kim}, \citenamefont {Oh},\ and\ \citenamefont
  {Kim}}]{jung2012stability}%
  \BibitemOpen
  \bibfield  {author} {\bibinfo {author} {\bibfnamefont {J.}~\bibnamefont
  {Jung}}, \bibinfo {author} {\bibfnamefont {D.~L.}\ \bibnamefont {Kim}},
  \bibinfo {author} {\bibfnamefont {S.~H.}\ \bibnamefont {Oh}}, \ and\ \bibinfo
  {author} {\bibfnamefont {H.~J.}\ \bibnamefont {Kim}},\ }\href@noop {}
  {\bibfield  {journal} {\bibinfo  {journal} {Solar energy materials and solar
  cells}\ }\textbf {\bibinfo {volume} {102}},\ \bibinfo {pages} {103} (\bibinfo
  {year} {2012})}\BibitemShut {NoStop}%
\bibitem [{\citenamefont {Hubbard}(1964)}]{hubbard1964electron}%
  \BibitemOpen
  \bibfield  {author} {\bibinfo {author} {\bibfnamefont {J.}~\bibnamefont
  {Hubbard}},\ }\href@noop {} {\emph {\bibinfo {title} {Proceedings of the
  Royal Society of London A: Mathematical, Physical and Engineering
  Sciences}}},\ Vol.\ \bibinfo {volume} {281}\ (\bibinfo  {publisher} {The
  Royal Society},\ \bibinfo {year} {1964})\ pp.\ \bibinfo {pages}
  {401--419}\BibitemShut {NoStop}%
\bibitem [{\citenamefont {R{\"o}dl}\ \emph {et~al.}(2009)\citenamefont
  {R{\"o}dl}, \citenamefont {Fuchs}, \citenamefont {Furthm{\"u}ller},\ and\
  \citenamefont {Bechstedt}}]{rodl2009quasiparticle}%
  \BibitemOpen
  \bibfield  {author} {\bibinfo {author} {\bibfnamefont {C.}~\bibnamefont
  {R{\"o}dl}}, \bibinfo {author} {\bibfnamefont {F.}~\bibnamefont {Fuchs}},
  \bibinfo {author} {\bibfnamefont {J.}~\bibnamefont {Furthm{\"u}ller}}, \ and\
  \bibinfo {author} {\bibfnamefont {F.}~\bibnamefont {Bechstedt}},\ }\href@noop
  {} {\bibfield  {journal} {\bibinfo  {journal} {Phys. Rev. B}\ }\textbf
  {\bibinfo {volume} {79}},\ \bibinfo {pages} {235114} (\bibinfo {year}
  {2009})}\BibitemShut {NoStop}%
\bibitem [{\citenamefont {Zhang}\ and\ \citenamefont
  {Rice}(1988)}]{zhang1988effective}%
  \BibitemOpen
  \bibfield  {author} {\bibinfo {author} {\bibfnamefont {F.}~\bibnamefont
  {Zhang}}\ and\ \bibinfo {author} {\bibfnamefont {T.}~\bibnamefont {Rice}},\
  }\href@noop {} {\bibfield  {journal} {\bibinfo  {journal} {Phys. Rev. B}\
  }\textbf {\bibinfo {volume} {37}},\ \bibinfo {pages} {3759} (\bibinfo {year}
  {1988})}\BibitemShut {NoStop}%
\bibitem [{\citenamefont {Kune{\v{s}}}\ \emph {et~al.}(2007)\citenamefont
  {Kune{\v{s}}}, \citenamefont {Anisimov}, \citenamefont {Skornyakov},
  \citenamefont {Lukoyanov},\ and\ \citenamefont {Vollhardt}}]{kunevs2007nio}%
  \BibitemOpen
  \bibfield  {author} {\bibinfo {author} {\bibfnamefont {J.}~\bibnamefont
  {Kune{\v{s}}}}, \bibinfo {author} {\bibfnamefont {V.}~\bibnamefont
  {Anisimov}}, \bibinfo {author} {\bibfnamefont {S.}~\bibnamefont
  {Skornyakov}}, \bibinfo {author} {\bibfnamefont {A.}~\bibnamefont
  {Lukoyanov}}, \ and\ \bibinfo {author} {\bibfnamefont {D.}~\bibnamefont
  {Vollhardt}},\ }\href@noop {} {\bibfield  {journal} {\bibinfo  {journal}
  {Phys. Rev. Lett.}\ }\textbf {\bibinfo {volume} {99}},\ \bibinfo {pages}
  {156404} (\bibinfo {year} {2007})}\BibitemShut {NoStop}%
\bibitem [{\citenamefont {Ba{\l}a}\ \emph {et~al.}(1994)\citenamefont
  {Ba{\l}a}, \citenamefont {Ole{\'s}},\ and\ \citenamefont
  {Zaanen}}]{bala1994zhang}%
  \BibitemOpen
  \bibfield  {author} {\bibinfo {author} {\bibfnamefont {J.}~\bibnamefont
  {Ba{\l}a}}, \bibinfo {author} {\bibfnamefont {A.~M.}\ \bibnamefont
  {Ole{\'s}}}, \ and\ \bibinfo {author} {\bibfnamefont {J.}~\bibnamefont
  {Zaanen}},\ }\href@noop {} {\bibfield  {journal} {\bibinfo  {journal} {Phys.
  Rev. Lett.}\ }\textbf {\bibinfo {volume} {72}},\ \bibinfo {pages} {2600}
  (\bibinfo {year} {1994})}\BibitemShut {NoStop}%
\bibitem [{\citenamefont {Damascelli}\ \emph {et~al.}(2003)\citenamefont
  {Damascelli}, \citenamefont {Hussain},\ and\ \citenamefont
  {Shen}}]{damascelli2003angle}%
  \BibitemOpen
  \bibfield  {author} {\bibinfo {author} {\bibfnamefont {A.}~\bibnamefont
  {Damascelli}}, \bibinfo {author} {\bibfnamefont {Z.}~\bibnamefont {Hussain}},
  \ and\ \bibinfo {author} {\bibfnamefont {Z.-X.}\ \bibnamefont {Shen}},\
  }\href@noop {} {\bibfield  {journal} {\bibinfo  {journal} {Reviews of modern
  physics}\ }\textbf {\bibinfo {volume} {75}},\ \bibinfo {pages} {473}
  (\bibinfo {year} {2003})}\BibitemShut {NoStop}%
\bibitem [{\citenamefont {Monney}\ \emph {et~al.}(2016)\citenamefont {Monney},
  \citenamefont {Bisogni}, \citenamefont {Zhou}, \citenamefont {Kraus},
  \citenamefont {Strocov}, \citenamefont {Behr}, \citenamefont {Drechsler},
  \citenamefont {Rosner}, \citenamefont {Johnston}, \citenamefont {Geck} \emph
  {et~al.}}]{monney2016probing}%
  \BibitemOpen
  \bibfield  {author} {\bibinfo {author} {\bibfnamefont {C.}~\bibnamefont
  {Monney}}, \bibinfo {author} {\bibfnamefont {V.}~\bibnamefont {Bisogni}},
  \bibinfo {author} {\bibfnamefont {K.-J.}\ \bibnamefont {Zhou}}, \bibinfo
  {author} {\bibfnamefont {R.}~\bibnamefont {Kraus}}, \bibinfo {author}
  {\bibfnamefont {V.~N.}\ \bibnamefont {Strocov}}, \bibinfo {author}
  {\bibfnamefont {G.}~\bibnamefont {Behr}}, \bibinfo {author} {\bibfnamefont
  {S.-L.}\ \bibnamefont {Drechsler}}, \bibinfo {author} {\bibfnamefont
  {H.}~\bibnamefont {Rosner}}, \bibinfo {author} {\bibfnamefont
  {S.}~\bibnamefont {Johnston}}, \bibinfo {author} {\bibfnamefont
  {J.}~\bibnamefont {Geck}},  \emph {et~al.},\ }\href@noop {} {\bibfield
  {journal} {\bibinfo  {journal} {Phys. Rev. B}\ }\textbf {\bibinfo {volume}
  {94}},\ \bibinfo {pages} {165118} (\bibinfo {year} {2016})}\BibitemShut
  {NoStop}%
\bibitem [{\citenamefont {Barman}\ \emph {et~al.}(2020)\citenamefont {Barman},
  \citenamefont {Kundu},\ and\ \citenamefont {Menon}}]{barman}%
  \BibitemOpen
  \bibfield  {author} {\bibinfo {author} {\bibfnamefont {S.}~\bibnamefont
  {Barman}}, \bibinfo {author} {\bibfnamefont {A.~K.}\ \bibnamefont {Kundu}}, \
  and\ \bibinfo {author} {\bibfnamefont {K.~S.~R.}\ \bibnamefont {Menon}},\
  }\href@noop {} {\bibfield  {journal} {\bibinfo  {journal} {Journal of
  Magnetism and Magnetic Materials}\ }\textbf {\bibinfo {volume} {515}},\
  \bibinfo {pages} {167292} (\bibinfo {year} {2020})}\BibitemShut {NoStop}%
\bibitem [{\citenamefont {Hermsmeier}\ \emph {et~al.}(1990)\citenamefont
  {Hermsmeier}, \citenamefont {Osterwalder}, \citenamefont {Friedman},
  \citenamefont {Sinkovic}, \citenamefont {Tran},\ and\ \citenamefont
  {Fadley}}]{hermsmeier1990spin}%
  \BibitemOpen
  \bibfield  {author} {\bibinfo {author} {\bibfnamefont {B.}~\bibnamefont
  {Hermsmeier}}, \bibinfo {author} {\bibfnamefont {J.}~\bibnamefont
  {Osterwalder}}, \bibinfo {author} {\bibfnamefont {D.}~\bibnamefont
  {Friedman}}, \bibinfo {author} {\bibfnamefont {B.}~\bibnamefont {Sinkovic}},
  \bibinfo {author} {\bibfnamefont {T.}~\bibnamefont {Tran}}, \ and\ \bibinfo
  {author} {\bibfnamefont {C.}~\bibnamefont {Fadley}},\ }\href@noop {}
  {\bibfield  {journal} {\bibinfo  {journal} {Phys. Rev. B}\ }\textbf {\bibinfo
  {volume} {42}},\ \bibinfo {pages} {11895} (\bibinfo {year}
  {1990})}\BibitemShut {NoStop}%
\bibitem [{\citenamefont {Kundu}\ \emph {et~al.}(2017)\citenamefont {Kundu},
  \citenamefont {Barman},\ and\ \citenamefont {Menon}}]{kundu2017effects}%
  \BibitemOpen
  \bibfield  {author} {\bibinfo {author} {\bibfnamefont {A.~K.}\ \bibnamefont
  {Kundu}}, \bibinfo {author} {\bibfnamefont {S.}~\bibnamefont {Barman}}, \
  and\ \bibinfo {author} {\bibfnamefont {K.~S.~R.}\ \bibnamefont {Menon}},\
  }\href@noop {} {\bibfield  {journal} {\bibinfo  {journal} {Phys. Rev. B}\
  }\textbf {\bibinfo {volume} {96}},\ \bibinfo {pages} {195116} (\bibinfo
  {year} {2017})}\BibitemShut {NoStop}%
\bibitem [{\citenamefont {Barman}\ and\ \citenamefont
  {Menon}(2018)}]{barman2018growth}%
  \BibitemOpen
  \bibfield  {author} {\bibinfo {author} {\bibfnamefont {S.}~\bibnamefont
  {Barman}}\ and\ \bibinfo {author} {\bibfnamefont {K.~S.~R.}\ \bibnamefont
  {Menon}},\ }\href@noop {} {\bibfield  {journal} {\bibinfo  {journal} {Journal
  of Crystal Growth}\ }\textbf {\bibinfo {volume} {487}},\ \bibinfo {pages}
  {28} (\bibinfo {year} {2018})}\BibitemShut {NoStop}%
\bibitem [{\citenamefont {Kundu}\ and\ \citenamefont
  {Menon}(2016)}]{kundu2016growth}%
  \BibitemOpen
  \bibfield  {author} {\bibinfo {author} {\bibfnamefont {A.~K.}\ \bibnamefont
  {Kundu}}\ and\ \bibinfo {author} {\bibfnamefont {K.~S.~R.}\ \bibnamefont
  {Menon}},\ }\href@noop {} {\bibfield  {journal} {\bibinfo  {journal} {Journal
  of Crystal Growth}\ }\textbf {\bibinfo {volume} {446}},\ \bibinfo {pages}
  {85} (\bibinfo {year} {2016})}\BibitemShut {NoStop}%
\bibitem [{\citenamefont {Das}\ and\ \citenamefont
  {Menon}(2015)}]{das2015revisit}%
  \BibitemOpen
  \bibfield  {author} {\bibinfo {author} {\bibfnamefont {J.}~\bibnamefont
  {Das}}\ and\ \bibinfo {author} {\bibfnamefont {K.~S.~R.}\ \bibnamefont
  {Menon}},\ }\href@noop {} {\bibfield  {journal} {\bibinfo  {journal} {Journal
  of Electron Spectroscopy and Related Phenomena}\ }\textbf {\bibinfo {volume}
  {203}},\ \bibinfo {pages} {71} (\bibinfo {year} {2015})}\BibitemShut
  {NoStop}%
\bibitem [{\citenamefont {Terakura}\ \emph {et~al.}(1984)\citenamefont
  {Terakura}, \citenamefont {Oguchi}, \citenamefont {Williams},\ and\
  \citenamefont {K{\"u}bler}}]{terakura1984band}%
  \BibitemOpen
  \bibfield  {author} {\bibinfo {author} {\bibfnamefont {K.}~\bibnamefont
  {Terakura}}, \bibinfo {author} {\bibfnamefont {T.}~\bibnamefont {Oguchi}},
  \bibinfo {author} {\bibfnamefont {A.}~\bibnamefont {Williams}}, \ and\
  \bibinfo {author} {\bibfnamefont {J.}~\bibnamefont {K{\"u}bler}},\
  }\href@noop {} {\bibfield  {journal} {\bibinfo  {journal} {Phys. Rev. B}\
  }\textbf {\bibinfo {volume} {30}},\ \bibinfo {pages} {4734} (\bibinfo {year}
  {1984})}\BibitemShut {NoStop}%
\bibitem [{\citenamefont {Shen}\ \emph
  {et~al.}(1990{\natexlab{b}})\citenamefont {Shen}, \citenamefont {Shih},
  \citenamefont {Jepsen}, \citenamefont {Spicer}, \citenamefont {Lindau},\ and\
  \citenamefont {Allen}}]{shen1990aspects}%
  \BibitemOpen
  \bibfield  {author} {\bibinfo {author} {\bibfnamefont {Z.-X.}\ \bibnamefont
  {Shen}}, \bibinfo {author} {\bibfnamefont {C.}~\bibnamefont {Shih}}, \bibinfo
  {author} {\bibfnamefont {O.}~\bibnamefont {Jepsen}}, \bibinfo {author}
  {\bibfnamefont {W.}~\bibnamefont {Spicer}}, \bibinfo {author} {\bibfnamefont
  {I.}~\bibnamefont {Lindau}}, \ and\ \bibinfo {author} {\bibfnamefont
  {J.}~\bibnamefont {Allen}},\ }\href@noop {} {\bibfield  {journal} {\bibinfo
  {journal} {Phys. Rev. Lett.}\ }\textbf {\bibinfo {volume} {64}},\ \bibinfo
  {pages} {2442} (\bibinfo {year} {1990}{\natexlab{b}})}\BibitemShut {NoStop}%
\bibitem [{\citenamefont {Mandal}\ \emph
  {et~al.}(2019{\natexlab{b}})\citenamefont {Mandal}, \citenamefont {Haule},
  \citenamefont {Rabe},\ and\ \citenamefont
  {Vanderbilt}}]{mandal2019systematic}%
  \BibitemOpen
  \bibfield  {author} {\bibinfo {author} {\bibfnamefont {S.}~\bibnamefont
  {Mandal}}, \bibinfo {author} {\bibfnamefont {K.}~\bibnamefont {Haule}},
  \bibinfo {author} {\bibfnamefont {K.~M.}\ \bibnamefont {Rabe}}, \ and\
  \bibinfo {author} {\bibfnamefont {D.}~\bibnamefont {Vanderbilt}},\
  }\href@noop {} {\bibfield  {journal} {\bibinfo  {journal} {npj Computational
  Materials}\ }\textbf {\bibinfo {volume} {5}},\ \bibinfo {pages} {1} (\bibinfo
  {year} {2019}{\natexlab{b}})}\BibitemShut {NoStop}%
\bibitem [{\citenamefont {Yin}\ \emph {et~al.}(2011)\citenamefont {Yin},
  \citenamefont {Haule},\ and\ \citenamefont {Kotliar}}]{haule3}%
  \BibitemOpen
  \bibfield  {author} {\bibinfo {author} {\bibfnamefont {Z.~P.}\ \bibnamefont
  {Yin}}, \bibinfo {author} {\bibfnamefont {K.}~\bibnamefont {Haule}}, \ and\
  \bibinfo {author} {\bibfnamefont {G.}~\bibnamefont {Kotliar}},\ }\href@noop
  {} {\bibfield  {journal} {\bibinfo  {journal} {Nat. Mater.}\ }\textbf
  {\bibinfo {volume} {10}},\ \bibinfo {pages} {932} (\bibinfo {year}
  {2011})}\BibitemShut {NoStop}%
\bibitem [{\citenamefont {Haule}\ \emph {et~al.}(2010)\citenamefont {Haule},
  \citenamefont {Yee},\ and\ \citenamefont {Kim}}]{Haule_prb10}%
  \BibitemOpen
  \bibfield  {author} {\bibinfo {author} {\bibfnamefont {K.}~\bibnamefont
  {Haule}}, \bibinfo {author} {\bibfnamefont {C.-H.}\ \bibnamefont {Yee}}, \
  and\ \bibinfo {author} {\bibfnamefont {K.}~\bibnamefont {Kim}},\ }\href
  {\doibase 10.1103/PhysRevB.81.195107} {\bibfield  {journal} {\bibinfo
  {journal} {Phys. Rev. B}\ }\textbf {\bibinfo {volume} {81}},\ \bibinfo
  {pages} {195107} (\bibinfo {year} {2010})}\BibitemShut {NoStop}%
\bibitem [{\citenamefont {Kotliar}\ \emph {et~al.}(2006)\citenamefont
  {Kotliar}, \citenamefont {Savrasov}, \citenamefont {Haule}, \citenamefont
  {Oudovenko}, \citenamefont {Parcollet},\ and\ \citenamefont
  {Marianetti}}]{DMFT_review}%
  \BibitemOpen
  \bibfield  {author} {\bibinfo {author} {\bibfnamefont {G.}~\bibnamefont
  {Kotliar}}, \bibinfo {author} {\bibfnamefont {S.~Y.}\ \bibnamefont
  {Savrasov}}, \bibinfo {author} {\bibfnamefont {K.}~\bibnamefont {Haule}},
  \bibinfo {author} {\bibfnamefont {V.~S.}\ \bibnamefont {Oudovenko}}, \bibinfo
  {author} {\bibfnamefont {O.}~\bibnamefont {Parcollet}}, \ and\ \bibinfo
  {author} {\bibfnamefont {C.~A.}\ \bibnamefont {Marianetti}},\ }\href
  {\doibase 10.1103/RevModPhys.78.865} {\bibfield  {journal} {\bibinfo
  {journal} {Rev. Mod. Phys.}\ }\textbf {\bibinfo {volume} {78}},\ \bibinfo
  {pages} {865} (\bibinfo {year} {2006})}\BibitemShut {NoStop}%
\bibitem [{\citenamefont {Kune{\v s}}\ \emph {et~al.}(2008)\citenamefont
  {Kune{\v s}}, \citenamefont {Yamasaki}, \citenamefont {Lukoyanov},
  \citenamefont {Feldbacher}, \citenamefont {Anisimov}, \citenamefont {Yang},
  \citenamefont {Scalettar}, \citenamefont {Andersen}, \citenamefont
  {Pickett},\ and\ \citenamefont {Held}}]{Kunes:2008bh}%
  \BibitemOpen
  \bibfield  {author} {\bibinfo {author} {\bibfnamefont {J.}~\bibnamefont
  {Kune{\v s}}}, \bibinfo {author} {\bibfnamefont {A.}~\bibnamefont
  {Yamasaki}}, \bibinfo {author} {\bibfnamefont {A.~V.}\ \bibnamefont
  {Lukoyanov}}, \bibinfo {author} {\bibfnamefont {M.}~\bibnamefont
  {Feldbacher}}, \bibinfo {author} {\bibfnamefont {V.~I.}\ \bibnamefont
  {Anisimov}}, \bibinfo {author} {\bibfnamefont {Y.~F.}\ \bibnamefont {Yang}},
  \bibinfo {author} {\bibfnamefont {R.~T.}\ \bibnamefont {Scalettar}}, \bibinfo
  {author} {\bibfnamefont {O.}~\bibnamefont {Andersen}}, \bibinfo {author}
  {\bibfnamefont {W.~E.}\ \bibnamefont {Pickett}}, \ and\ \bibinfo {author}
  {\bibfnamefont {K.}~\bibnamefont {Held}},\ }\href@noop {} {\bibfield
  {journal} {\bibinfo  {journal} {Nature Mat.}\ }\textbf {\bibinfo {volume}
  {7}},\ \bibinfo {pages} {198} (\bibinfo {year} {2008})}\BibitemShut {NoStop}%
\bibitem [{\citenamefont {Mandal}\ \emph {et~al.}(2017)\citenamefont {Mandal},
  \citenamefont {Zhang}, \citenamefont {Ismail-Beigi},\ and\ \citenamefont
  {Haule}}]{FeSe_monolayer}%
  \BibitemOpen
  \bibfield  {author} {\bibinfo {author} {\bibfnamefont {S.}~\bibnamefont
  {Mandal}}, \bibinfo {author} {\bibfnamefont {P.}~\bibnamefont {Zhang}},
  \bibinfo {author} {\bibfnamefont {S.}~\bibnamefont {Ismail-Beigi}}, \ and\
  \bibinfo {author} {\bibfnamefont {K.}~\bibnamefont {Haule}},\ }\href
  {https://journals.aps.org/prl/pdf/10.1103/PhysRevLett.119.067004} {\bibfield
  {journal} {\bibinfo  {journal} {Phys. Rev. Lett.}\ }\textbf {\bibinfo
  {volume} {119}},\ \bibinfo {pages} {067004} (\bibinfo {year}
  {2017})}\BibitemShut {NoStop}%
\bibitem [{\citenamefont {Mandal}\ \emph
  {et~al.}(2014{\natexlab{a}})\citenamefont {Mandal}, \citenamefont {Cohen},\
  and\ \citenamefont {Haule}}]{Mandal:2014}%
  \BibitemOpen
  \bibfield  {author} {\bibinfo {author} {\bibfnamefont {S.}~\bibnamefont
  {Mandal}}, \bibinfo {author} {\bibfnamefont {R.~E.}\ \bibnamefont {Cohen}}, \
  and\ \bibinfo {author} {\bibfnamefont {K.}~\bibnamefont {Haule}},\ }\href
  {https://journals.aps.org/prb/abstract/10.1103/PhysRevB.90.060501} {\bibfield
   {journal} {\bibinfo  {journal} {Phys. Rev. B}\ }\textbf {\bibinfo {volume}
  {90}},\ \bibinfo {pages} {060501} (\bibinfo {year}
  {2014}{\natexlab{a}})}\BibitemShut {NoStop}%
\bibitem [{\citenamefont {Mandal}\ \emph
  {et~al.}(2014{\natexlab{b}})\citenamefont {Mandal}, \citenamefont {Cohen},\
  and\ \citenamefont {Haule}}]{Mandal2:2014}%
  \BibitemOpen
  \bibfield  {author} {\bibinfo {author} {\bibfnamefont {S.}~\bibnamefont
  {Mandal}}, \bibinfo {author} {\bibfnamefont {R.~E.}\ \bibnamefont {Cohen}}, \
  and\ \bibinfo {author} {\bibfnamefont {K.}~\bibnamefont {Haule}},\ }\href
  {https://journals.aps.org/prb/abstract/10.1103/PhysRevB.89.220502} {\bibfield
   {journal} {\bibinfo  {journal} {Phys. Rev. B}\ }\textbf {\bibinfo {volume}
  {89}},\ \bibinfo {pages} {220502} (\bibinfo {year}
  {2014}{\natexlab{b}})}\BibitemShut {NoStop}%
\bibitem [{\citenamefont {Mandal}\ \emph {et~al.}(2018)\citenamefont {Mandal},
  \citenamefont {Cohen},\ and\ \citenamefont {Haule}}]{Mandal:2018}%
  \BibitemOpen
  \bibfield  {author} {\bibinfo {author} {\bibfnamefont {S.}~\bibnamefont
  {Mandal}}, \bibinfo {author} {\bibfnamefont {R.~E.}\ \bibnamefont {Cohen}}, \
  and\ \bibinfo {author} {\bibfnamefont {K.}~\bibnamefont {Haule}},\ }\href
  {\doibase 10.1103/PhysRevB.98.075155} {\bibfield  {journal} {\bibinfo
  {journal} {Phys. Rev. B}\ }\textbf {\bibinfo {volume} {98}},\ \bibinfo
  {pages} {075155} (\bibinfo {year} {2018})}\BibitemShut {NoStop}%
\bibitem [{\citenamefont {Hariki}\ \emph {et~al.}(2013)\citenamefont {Hariki},
  \citenamefont {Ichinozuka},\ and\ \citenamefont
  {Uozumi}}]{hariki2013dynamical}%
  \BibitemOpen
  \bibfield  {author} {\bibinfo {author} {\bibfnamefont {A.}~\bibnamefont
  {Hariki}}, \bibinfo {author} {\bibfnamefont {Y.}~\bibnamefont {Ichinozuka}},
  \ and\ \bibinfo {author} {\bibfnamefont {T.}~\bibnamefont {Uozumi}},\
  }\href@noop {} {\bibfield  {journal} {\bibinfo  {journal} {Journal of the
  Physical Society of Japan}\ }\textbf {\bibinfo {volume} {82}},\ \bibinfo
  {pages} {043710} (\bibinfo {year} {2013})}\BibitemShut {NoStop}%
\bibitem [{\citenamefont {Hariki}\ \emph {et~al.}(2017)\citenamefont {Hariki},
  \citenamefont {Uozumi},\ and\ \citenamefont {Kune\ifmmode~\check{s}\else
  \v{s}\fi{}}}]{hariki2017lda}%
  \BibitemOpen
  \bibfield  {author} {\bibinfo {author} {\bibfnamefont {A.}~\bibnamefont
  {Hariki}}, \bibinfo {author} {\bibfnamefont {T.}~\bibnamefont {Uozumi}}, \
  and\ \bibinfo {author} {\bibfnamefont {J.}~\bibnamefont
  {Kune\ifmmode~\check{s}\else \v{s}\fi{}}},\ }\href {\doibase
  10.1103/PhysRevB.96.045111} {\bibfield  {journal} {\bibinfo  {journal} {Phys.
  Rev. B}\ }\textbf {\bibinfo {volume} {96}},\ \bibinfo {pages} {045111}
  (\bibinfo {year} {2017})}\BibitemShut {NoStop}%
\bibitem [{\citenamefont {Blech}\ and\ \citenamefont
  {Averbach}(1966)}]{blech1966long}%
  \BibitemOpen
  \bibfield  {author} {\bibinfo {author} {\bibfnamefont {I.}~\bibnamefont
  {Blech}}\ and\ \bibinfo {author} {\bibfnamefont {B.}~\bibnamefont
  {Averbach}},\ }\href@noop {} {\bibfield  {journal} {\bibinfo  {journal}
  {Physical Review}\ }\textbf {\bibinfo {volume} {142}},\ \bibinfo {pages}
  {287} (\bibinfo {year} {1966})}\BibitemShut {NoStop}%
\bibitem [{\citenamefont {Kundu}\ \emph {et~al.}(2018)\citenamefont {Kundu},
  \citenamefont {Barman},\ and\ \citenamefont {Menon}}]{kundu2018evolution}%
  \BibitemOpen
  \bibfield  {author} {\bibinfo {author} {\bibfnamefont {A.~K.}\ \bibnamefont
  {Kundu}}, \bibinfo {author} {\bibfnamefont {S.}~\bibnamefont {Barman}}, \
  and\ \bibinfo {author} {\bibfnamefont {K.~S.~R.}\ \bibnamefont {Menon}},\
  }\href@noop {} {\bibfield  {journal} {\bibinfo  {journal} {Journal of
  Magnetism and Magnetic Materials}\ }\textbf {\bibinfo {volume} {466}},\
  \bibinfo {pages} {186} (\bibinfo {year} {2018})}\BibitemShut {NoStop}%
\bibitem [{\citenamefont {Menon}\ \emph {et~al.}(2011)\citenamefont {Menon},
  \citenamefont {Mandal}, \citenamefont {Das}, \citenamefont {Mente{\c{s}}},
  \citenamefont {Ni{\~n}o}, \citenamefont {Locatelli},\ and\ \citenamefont
  {Belkhou}}]{menon2011surface}%
  \BibitemOpen
  \bibfield  {author} {\bibinfo {author} {\bibfnamefont {K.~S.~R.}\
  \bibnamefont {Menon}}, \bibinfo {author} {\bibfnamefont {S.}~\bibnamefont
  {Mandal}}, \bibinfo {author} {\bibfnamefont {J.}~\bibnamefont {Das}},
  \bibinfo {author} {\bibfnamefont {T.~O.}\ \bibnamefont {Mente{\c{s}}}},
  \bibinfo {author} {\bibfnamefont {M.~A.}\ \bibnamefont {Ni{\~n}o}}, \bibinfo
  {author} {\bibfnamefont {A.}~\bibnamefont {Locatelli}}, \ and\ \bibinfo
  {author} {\bibfnamefont {R.}~\bibnamefont {Belkhou}},\ }\href@noop {}
  {\bibfield  {journal} {\bibinfo  {journal} {Phys. Rev. B}\ }\textbf {\bibinfo
  {volume} {84}},\ \bibinfo {pages} {132402} (\bibinfo {year}
  {2011})}\BibitemShut {NoStop}%
\bibitem [{\citenamefont {Das}\ and\ \citenamefont
  {Menon}(2018)}]{das2018evolution}%
  \BibitemOpen
  \bibfield  {author} {\bibinfo {author} {\bibfnamefont {J.}~\bibnamefont
  {Das}}\ and\ \bibinfo {author} {\bibfnamefont {K.~S.~R.}\ \bibnamefont
  {Menon}},\ }\href@noop {} {\bibfield  {journal} {\bibinfo  {journal} {Journal
  of Magnetism and Magnetic Materials}\ }\textbf {\bibinfo {volume} {449}},\
  \bibinfo {pages} {415} (\bibinfo {year} {2018})}\BibitemShut {NoStop}%
\bibitem [{\citenamefont {Lad}\ and\ \citenamefont
  {Henrich}(1988)}]{lad1988electronic}%
  \BibitemOpen
  \bibfield  {author} {\bibinfo {author} {\bibfnamefont {R.~J.}\ \bibnamefont
  {Lad}}\ and\ \bibinfo {author} {\bibfnamefont {V.~E.}\ \bibnamefont
  {Henrich}},\ }\href@noop {} {\bibfield  {journal} {\bibinfo  {journal} {Phys.
  Rev. B}\ }\textbf {\bibinfo {volume} {38}},\ \bibinfo {pages} {10860}
  (\bibinfo {year} {1988})}\BibitemShut {NoStop}%
\bibitem [{\citenamefont {Nekrasov}\ \emph {et~al.}(2013)\citenamefont
  {Nekrasov}, \citenamefont {Pavlov},\ and\ \citenamefont
  {Sadovskii}}]{nekrasov2013consistent}%
  \BibitemOpen
  \bibfield  {author} {\bibinfo {author} {\bibfnamefont {I.}~\bibnamefont
  {Nekrasov}}, \bibinfo {author} {\bibfnamefont {N.}~\bibnamefont {Pavlov}}, \
  and\ \bibinfo {author} {\bibfnamefont {M.}~\bibnamefont {Sadovskii}},\
  }\href@noop {} {\bibfield  {journal} {\bibinfo  {journal} {Journal of
  Experimental and Theoretical Physics}\ }\textbf {\bibinfo {volume} {116}},\
  \bibinfo {pages} {620} (\bibinfo {year} {2013})}\BibitemShut {NoStop}%
\bibitem [{\citenamefont {Eder}(2008)}]{eder2008correlated}%
  \BibitemOpen
  \bibfield  {author} {\bibinfo {author} {\bibfnamefont {R.}~\bibnamefont
  {Eder}},\ }\href@noop {} {\bibfield  {journal} {\bibinfo  {journal} {Phys.
  Rev. B}\ }\textbf {\bibinfo {volume} {78}},\ \bibinfo {pages} {115111}
  (\bibinfo {year} {2008})}\BibitemShut {NoStop}%
\bibitem [{\citenamefont {Trimarchi}\ \emph {et~al.}(2018)\citenamefont
  {Trimarchi}, \citenamefont {Wang},\ and\ \citenamefont
  {Zunger}}]{trimarchi2018polymorphous}%
  \BibitemOpen
  \bibfield  {author} {\bibinfo {author} {\bibfnamefont {G.}~\bibnamefont
  {Trimarchi}}, \bibinfo {author} {\bibfnamefont {Z.}~\bibnamefont {Wang}}, \
  and\ \bibinfo {author} {\bibfnamefont {A.}~\bibnamefont {Zunger}},\
  }\href@noop {} {\bibfield  {journal} {\bibinfo  {journal} {Phys. Rev. B}\
  }\textbf {\bibinfo {volume} {97}},\ \bibinfo {pages} {035107} (\bibinfo
  {year} {2018})}\BibitemShut {NoStop}%
\bibitem [{\citenamefont {Schmitt}\ \emph {et~al.}(2019)\citenamefont
  {Schmitt}, \citenamefont {Moras}, \citenamefont {Bihlmayer}, \citenamefont
  {Cotsakis}, \citenamefont {Vogt}, \citenamefont {Kemmer}, \citenamefont
  {Belabbes}, \citenamefont {Sheverdyaeva}, \citenamefont {Kundu},
  \citenamefont {Carbone} \emph {et~al.}}]{schmitt2019indirect}%
  \BibitemOpen
  \bibfield  {author} {\bibinfo {author} {\bibfnamefont {M.}~\bibnamefont
  {Schmitt}}, \bibinfo {author} {\bibfnamefont {P.}~\bibnamefont {Moras}},
  \bibinfo {author} {\bibfnamefont {G.}~\bibnamefont {Bihlmayer}}, \bibinfo
  {author} {\bibfnamefont {R.}~\bibnamefont {Cotsakis}}, \bibinfo {author}
  {\bibfnamefont {M.}~\bibnamefont {Vogt}}, \bibinfo {author} {\bibfnamefont
  {J.}~\bibnamefont {Kemmer}}, \bibinfo {author} {\bibfnamefont
  {A.}~\bibnamefont {Belabbes}}, \bibinfo {author} {\bibfnamefont {P.~M.}\
  \bibnamefont {Sheverdyaeva}}, \bibinfo {author} {\bibfnamefont {A.~K.}\
  \bibnamefont {Kundu}}, \bibinfo {author} {\bibfnamefont {C.}~\bibnamefont
  {Carbone}},  \emph {et~al.},\ }\href@noop {} {\bibfield  {journal} {\bibinfo
  {journal} {Nat Commun}\ }\textbf {\bibinfo {volume} {10}},\ \bibinfo {pages}
  {2610} (\bibinfo {year} {2019})}\BibitemShut {NoStop}%
\bibitem [{\citenamefont {Kundu}\ \emph {et~al.}(2021)\citenamefont {Kundu},
  \citenamefont {Barman},\ and\ \citenamefont {Menon}}]{kundu2021role}%
  \BibitemOpen
  \bibfield  {author} {\bibinfo {author} {\bibfnamefont {A.~K.}\ \bibnamefont
  {Kundu}}, \bibinfo {author} {\bibfnamefont {S.}~\bibnamefont {Barman}}, \
  and\ \bibinfo {author} {\bibfnamefont {K.~S.~R.}\ \bibnamefont {Menon}},\
  }\href@noop {} {\bibfield  {journal} {\bibinfo  {journal} {ACS Applied
  Materials \& Interfaces}\ }\textbf {\bibinfo {volume} {13}},\ \bibinfo
  {pages} {20779} (\bibinfo {year} {2021})}\BibitemShut {NoStop}%
\bibitem [{\citenamefont {Han}\ \emph {et~al.}(2023)\citenamefont {Han},
  \citenamefont {Telford}, \citenamefont {Kundu}, \citenamefont {Bintrim},
  \citenamefont {Turkel}, \citenamefont {Wiscons}, \citenamefont {Zangiabadi},
  \citenamefont {Choi}, \citenamefont {Li}, \citenamefont {Steigerwald} \emph
  {et~al.}}]{han2023interplay}%
  \BibitemOpen
  \bibfield  {author} {\bibinfo {author} {\bibfnamefont {S.~Y.}\ \bibnamefont
  {Han}}, \bibinfo {author} {\bibfnamefont {E.~J.}\ \bibnamefont {Telford}},
  \bibinfo {author} {\bibfnamefont {A.~K.}\ \bibnamefont {Kundu}}, \bibinfo
  {author} {\bibfnamefont {S.~J.}\ \bibnamefont {Bintrim}}, \bibinfo {author}
  {\bibfnamefont {S.}~\bibnamefont {Turkel}}, \bibinfo {author} {\bibfnamefont
  {R.~A.}\ \bibnamefont {Wiscons}}, \bibinfo {author} {\bibfnamefont
  {A.}~\bibnamefont {Zangiabadi}}, \bibinfo {author} {\bibfnamefont {E.-S.}\
  \bibnamefont {Choi}}, \bibinfo {author} {\bibfnamefont {T.-D.}\ \bibnamefont
  {Li}}, \bibinfo {author} {\bibfnamefont {M.~L.}\ \bibnamefont {Steigerwald}},
   \emph {et~al.},\ }\href@noop {} {\bibfield  {journal} {\bibinfo  {journal}
  {arXiv preprint arXiv:2307.01397}\ } (\bibinfo {year} {2023})}\BibitemShut
  {NoStop}%
\bibitem [{\citenamefont {Dagdeviren}\ \emph {et~al.}(2018)\citenamefont
  {Dagdeviren}, \citenamefont {Mandal}, \citenamefont {Zou}, \citenamefont
  {Zhou}, \citenamefont {Simon}, \citenamefont {Walker}, \citenamefont {Ahn},
  \citenamefont {Schwarz}, \citenamefont {Ismail-Beigi},\ and\ \citenamefont
  {Altman}}]{Topo-SM1}%
  \BibitemOpen
  \bibfield  {author} {\bibinfo {author} {\bibfnamefont {O.~E.}\ \bibnamefont
  {Dagdeviren}}, \bibinfo {author} {\bibfnamefont {S.}~\bibnamefont {Mandal}},
  \bibinfo {author} {\bibfnamefont {K.}~\bibnamefont {Zou}}, \bibinfo {author}
  {\bibfnamefont {C.}~\bibnamefont {Zhou}}, \bibinfo {author} {\bibfnamefont
  {G.~H.}\ \bibnamefont {Simon}}, \bibinfo {author} {\bibfnamefont {F.~J.}\
  \bibnamefont {Walker}}, \bibinfo {author} {\bibfnamefont {C.~H.}\
  \bibnamefont {Ahn}}, \bibinfo {author} {\bibfnamefont {U.~D.}\ \bibnamefont
  {Schwarz}}, \bibinfo {author} {\bibfnamefont {S.}~\bibnamefont
  {Ismail-Beigi}}, \ and\ \bibinfo {author} {\bibfnamefont {E.~I.}\
  \bibnamefont {Altman}},\ }\href {\doibase 10.1103/PhysRevMaterials.2.114205}
  {\bibfield  {journal} {\bibinfo  {journal} {Phys. Rev. Mater.}\ }\textbf
  {\bibinfo {volume} {2}},\ \bibinfo {pages} {114205} (\bibinfo {year}
  {2018})}\BibitemShut {NoStop}%
\bibitem [{\citenamefont {Renninger}\ \emph {et~al.}(1966)\citenamefont
  {Renninger}, \citenamefont {Moss},\ and\ \citenamefont {Averbach}}]{alan}%
  \BibitemOpen
  \bibfield  {author} {\bibinfo {author} {\bibfnamefont {A.}~\bibnamefont
  {Renninger}}, \bibinfo {author} {\bibfnamefont {S.}~\bibnamefont {Moss}}, \
  and\ \bibinfo {author} {\bibfnamefont {B.}~\bibnamefont {Averbach}},\
  }\href@noop {} {\bibfield  {journal} {\bibinfo  {journal} {Physical Review}\
  }\textbf {\bibinfo {volume} {147}},\ \bibinfo {pages} {418} (\bibinfo {year}
  {1966})}\BibitemShut {NoStop}%
\bibitem [{\citenamefont {Hermsmeier}\ \emph {et~al.}(1989)\citenamefont
  {Hermsmeier}, \citenamefont {Osterwalder}, \citenamefont {Friedman},\ and\
  \citenamefont {Fadley}}]{herm}%
  \BibitemOpen
  \bibfield  {author} {\bibinfo {author} {\bibfnamefont {B.}~\bibnamefont
  {Hermsmeier}}, \bibinfo {author} {\bibfnamefont {J.}~\bibnamefont
  {Osterwalder}}, \bibinfo {author} {\bibfnamefont {D.}~\bibnamefont
  {Friedman}}, \ and\ \bibinfo {author} {\bibfnamefont {C.}~\bibnamefont
  {Fadley}},\ }\href@noop {} {\bibfield  {journal} {\bibinfo  {journal} {Phys.
  Rev. lett.}\ }\textbf {\bibinfo {volume} {62}},\ \bibinfo {pages} {478}
  (\bibinfo {year} {1989})}\BibitemShut {NoStop}%
\bibitem [{\citenamefont {Horiba}\ \emph {et~al.}(2004)\citenamefont {Horiba},
  \citenamefont {Taguchi}, \citenamefont {Chainani}, \citenamefont {Takata},
  \citenamefont {Ikenaga}, \citenamefont {Miwa}, \citenamefont {Nishino},
  \citenamefont {Tamasaku}, \citenamefont {Awaji}, \citenamefont {Takeuchi}
  \emph {et~al.}}]{horiba2004nature}%
  \BibitemOpen
  \bibfield  {author} {\bibinfo {author} {\bibfnamefont {K.}~\bibnamefont
  {Horiba}}, \bibinfo {author} {\bibfnamefont {M.}~\bibnamefont {Taguchi}},
  \bibinfo {author} {\bibfnamefont {A.}~\bibnamefont {Chainani}}, \bibinfo
  {author} {\bibfnamefont {Y.}~\bibnamefont {Takata}}, \bibinfo {author}
  {\bibfnamefont {E.}~\bibnamefont {Ikenaga}}, \bibinfo {author} {\bibfnamefont
  {D.}~\bibnamefont {Miwa}}, \bibinfo {author} {\bibfnamefont {Y.}~\bibnamefont
  {Nishino}}, \bibinfo {author} {\bibfnamefont {K.}~\bibnamefont {Tamasaku}},
  \bibinfo {author} {\bibfnamefont {M.}~\bibnamefont {Awaji}}, \bibinfo
  {author} {\bibfnamefont {A.}~\bibnamefont {Takeuchi}},  \emph {et~al.},\
  }\href@noop {} {\bibfield  {journal} {\bibinfo  {journal} {Phys. Rev. Lett.}\
  }\textbf {\bibinfo {volume} {93}},\ \bibinfo {pages} {236401} (\bibinfo
  {year} {2004})}\BibitemShut {NoStop}%
\bibitem [{\citenamefont {van Veenendaal}(2006)}]{van2006competition}%
  \BibitemOpen
  \bibfield  {author} {\bibinfo {author} {\bibfnamefont {M.}~\bibnamefont {van
  Veenendaal}},\ }\href@noop {} {\bibfield  {journal} {\bibinfo  {journal}
  {Phys. Rev. B}\ }\textbf {\bibinfo {volume} {74}},\ \bibinfo {pages} {085118}
  (\bibinfo {year} {2006})}\BibitemShut {NoStop}%
\end{thebibliography}
%

\end{document}